\documentclass[preprint,5p]{elsarticle}

\usepackage{epsfig}

\usepackage{graphicx,xcolor}
\usepackage{epstopdf}

\usepackage{units,amsmath,amssymb,bm}
\usepackage{subfigure}

\usepackage{mathtools}
\usepackage{amsmath}
\usepackage{amssymb}
\usepackage{stmaryrd}
\usepackage{amsthm}
\usepackage{amsfonts}
\usepackage[utf8]{inputenc}
\usepackage[english]{babel}
\usepackage{multicol}
\usepackage{float}
\usepackage{textcomp}
\usepackage{gensymb}
\usepackage{mathrsfs}

\newcommand{\be}{\begin{equation}}
	\newcommand{\ee}{\end{equation}}

\newcommand{\rad}{\sqrt{2}}

\newcommand{\lcr}{\lambda_{\rm cr}}

\usepackage[normalem]{ulem} 
\newcommand\hl{\bgroup\markoverwith
	{\textcolor{yellow}{\rule[-.5ex]{2pt}{2.5ex}}}\ULon}

\journal{Fire And Materials}

\RequirePackage{lineno}

\begin{document}
	
	\begin{frontmatter}

\title{Thermomechanical surface instability at the origin of surface fissure patterns on heated circular MDF samples}

\author[tallinn]{Andrea Ferrantelli}


\author[aalto]{Djebar Baroudi}

\author[aalto,vtt]{Sergei Khakalo}

 \author[wuhan]{Kai Yuan Li\corref{cor1}}

 \cortext[cor1]{Corresponding author}
\ead{stevelikai@126.com}

\address[tallinn]{Tallinn University of Technology, Department of Civil Engineering and Architecture, 19086 Tallinn, Estonia}

\address[aalto]{Aalto University, Department of Civil Engineering, P.O.Box 12100, 00076 Aalto, Finland}

\address[vtt]{VTT Technical Research Centre of Finland, P.O.Box 1000, Espoo, Finland}

\address[wuhan]{School of Safety Science and Emergency Management, Wuhan University of Technology, Luoshi Road 122, Wuhan 430070, China}

\begin{abstract}

When a flat sample of medium density fibreboard (MDF) is exposed to radiant heat in an inert atmosphere, primary crack patterns suddenly start to appear over the entire surface before pyrolysis and any charring occurs. Contrary to common belief that crack formation is due to drying and shrinkage, it was demonstrated for square samples that this results from thermomechanical instability.

In the present paper, new experimental data are presented for circular samples of the same MDF material. The sample was exposed to radiant heating at 20 or 50 kW/m2, and completely different crack patterns with independent Eigenmodes were observed at the two heat fluxes.
We show that the two patterns can be reproduced with a full 3-D thermomechanical surface instability model of a hot layer adhered to an elastic colder foundation in an axisymmetric domain.  Analytical and numerical solutions of a simplified 2-D formulation of the same problem provide excellent qualitative agreement between observed and calculated patterns.

Previous data for square samples together with the results reported in the present paper for circular samples confirm the validity of the model for qualitative predictions, and indicate that further refinements can be made to improve its quantitative predictive capability.

\end{abstract}

\date{\today}

\end{frontmatter}

\section{Introduction}

When wood is heated, a well-known first effect that can be observed with rising temperature is the global and sudden appearance of regular crack patterns on its surface. These constitute preferred paths for the generation of new flames, and enhance the overall combustion and charring processes once the pyrolysis temperature ($\approx 300 ^\circ{\mathrm C}$) is reached. For these reasons, engineers preoccupied with fire safety have studied the crack patterns formation extensively \cite{ROBERTS1971893}.

Explanation was searched for a long time in physicochemical processes such as charring, drying and shrinkage, which occur at temperatures above the pyrolysis point \cite{Babrauskas2005528,Ross}. However, such previous investigations have not been able to physically explain the cracks pattern topology \cite{doi:10.1080/10618562.2012.659663,Stoliarov20102024,Lautenberger20091503}.

An alternative explanation might have been found by the authors of \cite{Baroudi2017206}, who investigated the physics of processes at temperatures \textit{below} the pyrolysis point, i.e. before any actual charring. In that study, the topology of patterns
 observed on oven-dry square wood and MDF samples, heated in nitrogen atmosphere, could be reproduced by assuming a thermomechanical surface instability that induces wrinkling.
In other words, under these conditions the principal crack pattern formation can be explained without considering any chemical processes.

Physical explanation is to be found indeed in the thermomechanical properties of wood (and its engineered byproduct MDF), which is a natural thermoplastic \cite{Salmen,Antoniow2012}: when reaching the glass transition temperature $T_g\approx200^oC$, dry wood simultaneously softens and elongates extensively. This induces in the hot layer restrain thermal stresses, which under certain conditions eventually lead to wrinkling  \cite{Baroudi2017206,Bazant_1985,Salmen_1984}.
As detailed in \cite{Baroudi2017206}, the cracks appear along the node lines of the buckling modes. This happens because there always exists a major principal tensile stress that is perpendicular to these nodes; the cracks initiate at locations where the mechanical resistance of the material decreases with increasing temperature.

The above mechanism has proven to be effective in reproducing the fissure patterns for \textit{square} wood and MDF samples \cite{Baroudi2017206}. In this paper we examine whether this is not restricted to that specific sample shape, namely if this novel explanation is general enough to apply to the topology observed on \textit{circular} MDF specimens as well.

To this aim, we discuss experiments performed on round oven-dry MDF samples that are heated from above in nitrogen atmosphere, and postulate a thermomechanical origin of the observed crack patterns as in \cite{Baroudi2017206}.
We try to match the observed crack topology by formulating a buckling model of a circular plate bonded to an elastic foundation,
successfully reproducing the observed patterns with a full 3-D model of thermomechanical buckling, which is solved numerically via Finite Element Method (FEM). Such model is formulated also in 2-D, both analytically and numerically, then validated against the literature \cite{Wang2004}.

\begin{figure}[t]
	\centering
	
	\begin{minipage}[b]{0.48\linewidth}
		
		\centering
		\includegraphics[width=\textwidth]{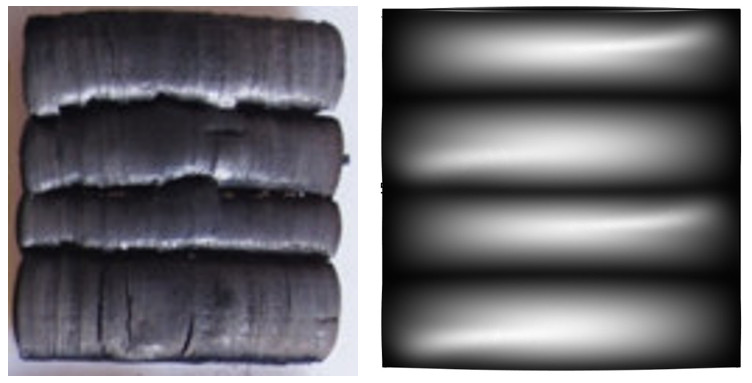}
		\caption{Fissure patterns on fir wood (orthotropic). Left: experiment, right: simulated \cite{Baroudi2017206}.}
		\label{fig:firpatterns}
	\end{minipage}
	\hspace{0.1cm}
	\begin{minipage}[b]{0.48\linewidth}
		\centering
		\includegraphics[width=\textwidth]{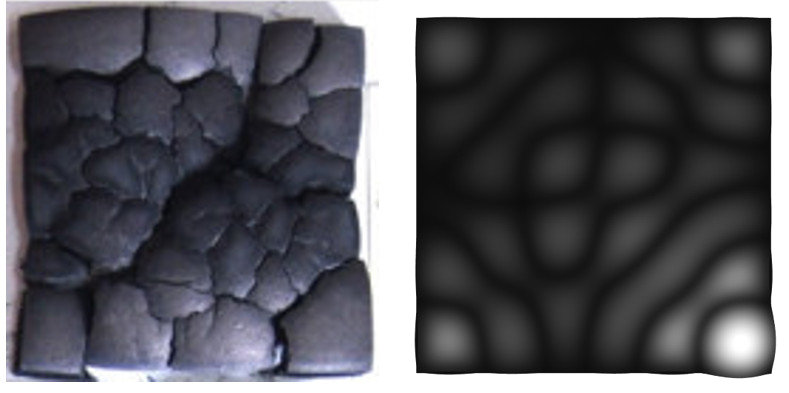}
		\caption{Fissure patterns on MDF wood (isotropic). Left: experiment, right: simulated \cite{Baroudi2017206}.}
		\label{fig:mdfpatterns}
	\end{minipage}
\end{figure}

 In conclusion, we argue that the same model of thermomechanical surface instability can explain the very different crack patterns emerging on both square (Figures \ref{fig:firpatterns} and \ref{fig:mdfpatterns}) and circular specimens, Figs.\ref{fig:20kW} and \ref{fig:50kW}.

Let us remark that we are focusing on soft matter, i.e. on the rubbery state of the heated layer prior to charring or burning, when thermal decomposition is still negligible. Moreover, the full mechanism of cracking formation might require further analysis: for instance, we do not investigate the crack initiation and propagation.

However, we know for sure that cracks appear at locations with maximal tensile stress; in this paper we limit ourselves to explaining the physical origin and topology of the observed crack patterns.
We should stress that the study performed here is indeed \textit{qualitative}: we are only preoccupied with confirming that the same thermomechanical model can explain the cracking patterns observed on both square and circular MDF specimens. We will compute exactly where and how much a crack opens with a full quantitative analysis, which will be presented in a forthcoming publication.

The present article is organized as follows: in Section \ref{sec:experimental-analysis} we discuss the experimental setup and measurements, while in Section \ref{sec:a-full-3-D-model-of-thermomechanical-buckling-and-its-validation} the 3-D model is described into detail. This is then compared to the observations in Section \ref{sec:experiments-interpretation-by-the-thermomechanical-model}, and conclusions are given in Section \ref{sec:conclusions}.

 The Appendix contains an analytical formulation of a 2-D model for the buckling of a thin layer bonded to an elastic substrate in \ref{sec:approximate-2-D-analytical-solution}. Finally, the corresponding numerical solution is obtained via variational formulation in 
 \ref{sec:variational-formulation-and-isogeometric-analysis-of-a-2-D-stability-problem}.


\section{Experimental setup and observations}\label{sec:experimental-analysis}

\subsection{Experimental setup}\label{Experimental setup}

\begin{figure}[t]
	\centering
	
	\begin{minipage}[b]{0.48\linewidth}
		\centering
		\includegraphics[width=\textwidth]{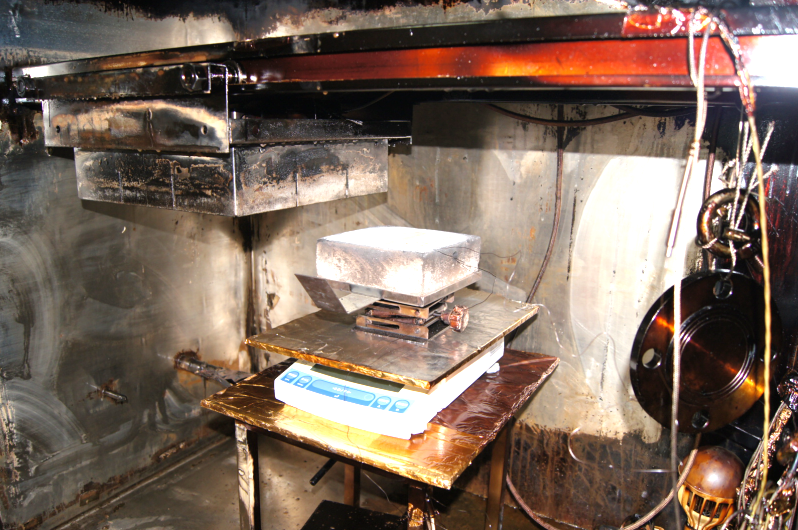}
		\caption{Internal view of the low pressure compartment.}
		\label{fig:compartment}

	\end{minipage}
	\hspace{0.1cm}
	\begin{minipage}[b]{0.48\linewidth}
		\centering
		\includegraphics[width=\textwidth]{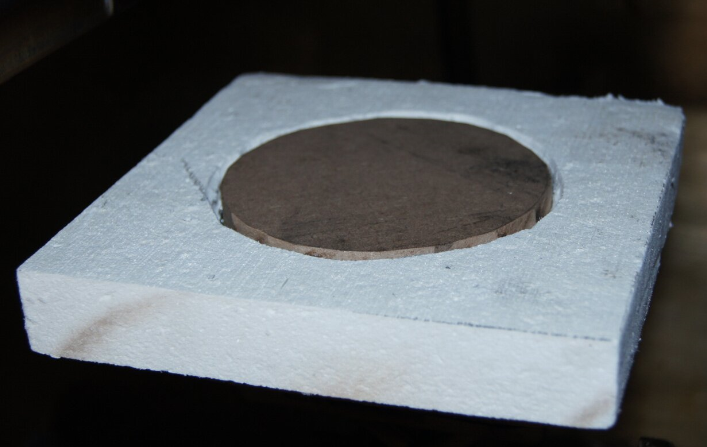}
		\caption{Sample into Kaowool holder.}
		\label{fig:kaowool}
		
	\end{minipage}
\end{figure}

The experimental rig consisted of the gas supply system (which provided nitrogen stored in a bottle and air from ambient environment through two pipes), a low pressure compartment (Figure \ref{fig:compartment}) and the control system.
A vacuum pump first reduced the absolute pressure in the compartment to 5 kPa, to remove most of the air. Pure nitrogen was then led in from the bottom of the compartment at a flowrate of 0.6 m$^3$/min. The vacuum pump was stopped until the pressure inside the chamber reached the targeting experimental pressure, and then turned on again to stabilize it.

Regarding the issue of preheating, this was not a problem for neither sample nor compartment temperatures. Our measurements show that for the sample, the surface temperature increased to around 50\textdegree C, hence much lower than the pyrolysis temperature, while the centre $T$ maintained the ambient value \cite{a3f4a276a46c4ecaa357a2fd51e95882}. The surface temperature rise was mainly caused by the radiation from the panel, which has no significant impact on the gas temperature inside the compartment (according to measurements, this increased by less than 10\textdegree C).

The samples were weighed via an electric balance, and two thermocouples measured the surface and internal temperatures of the sample. The surface thermocouple was freely attached to the sample surface, so that once the surface descended due to shrinkage, the thermocouple could follow while still attached to the sample surface. The centre thermocouple was placed by drilling a hole, penetrating the sample holder to reach the internal centre of the sample.
 A digital camera in front of the observation window was recording the experiment (more details and schemes are given in \cite{Baroudi2017206}).

The tested samples were produced by cutting a large piece of MDF board into several circular pieces with a diameter of 100 mm, thickness 15 mm and density 730 $\pm 17$ kg/m$^3$. A sample holder made of Kaowool with high heat insulation was used to carry the samples, as shown in Fig.\ref{fig:kaowool}. To identify the effect of heat flux and compare it to the previous patterns observed by \cite{Baroudi2017206} for rectangular samples, we used 20 kW/m$^2$ representing low heat flux and 50 kW/m$^2$ denoting high heat flux. The experiments were all performed under 95 kPa, which is close to the regular atmospheric pressure.

Finally, we used nitrogen atmosphere in order to prevent surface oxidation reactions that, after charring, would transform it into ash. The current setup simplifies the problem indeed, since using air will definitely lead to flaming condition in the current set of experiments. Adding flame and char oxidization would complicate the problem, e.g. the flame would add some extra heat flux that still constitutes an unsolved problem \cite{LI201539}.

For calculating the temperature and density profiles inside the samples, we used the pyrolysis model in Fire Dynamics Simulator (FDS) version 6.3.2 \cite{mcgrattan2013fire}. The model solves the coupled heat conduction and pyrolysis reaction equations with a one-dimensional finite difference analysis, summarized in \cite{Baroudi2017206}.

\subsection{Observations}\label{sec:experimental-observations}

\begin{figure}[t]
		
			
			\centering
			\includegraphics[width=0.48\textwidth]{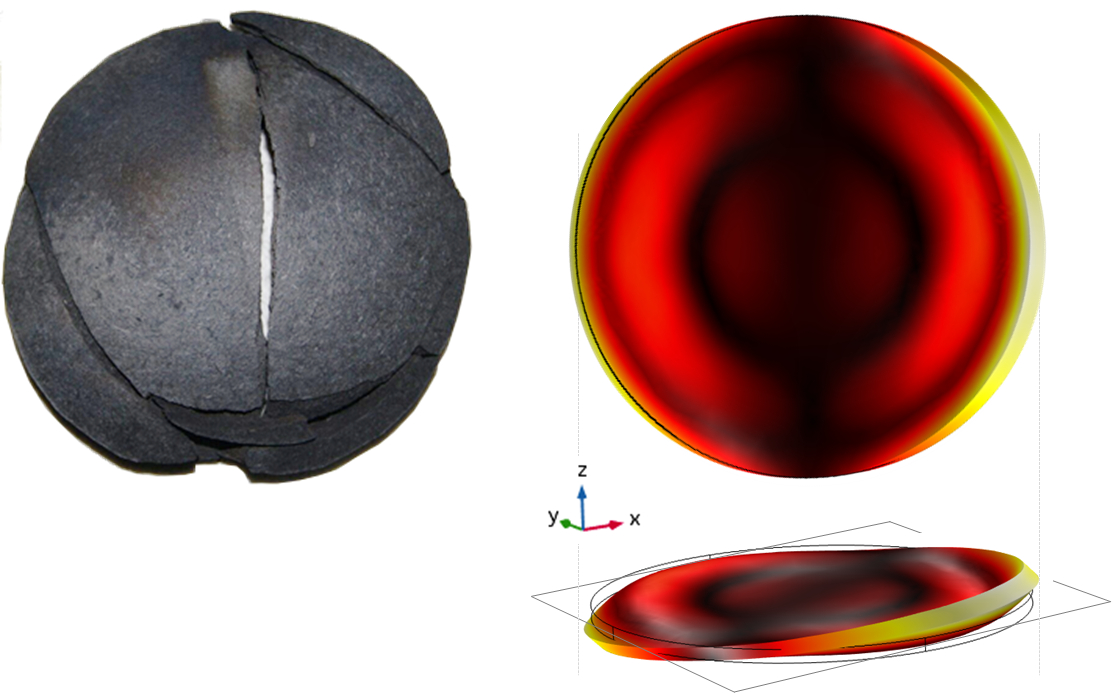}
			\caption{Fissure patterns for 20kW/m$^2$. Left: experiment, right: simulated (out-of-plane displacement in grayscale).}
			\label{fig:20kW}
\end{figure}

\begin{figure}
	\centering
	\includegraphics[width=0.48\textwidth]{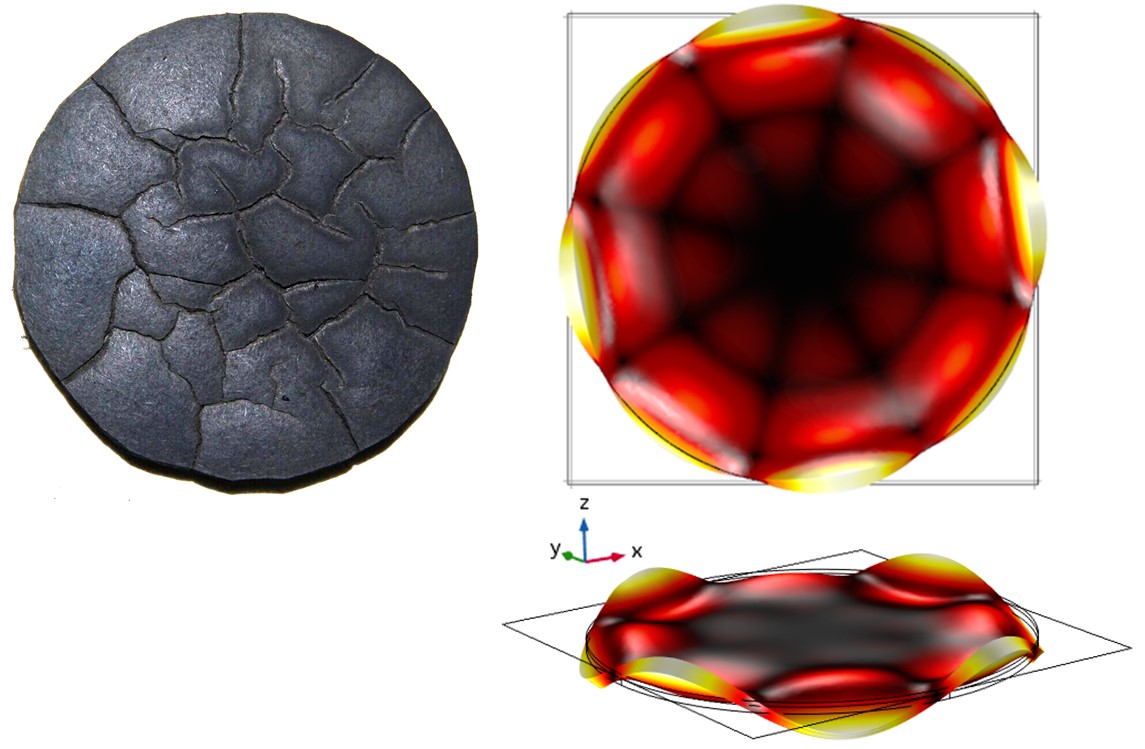}
	\caption{Fissure patterns for 50kW/m$^2$. Left: experiment, right: simulated (out-of-plane displacement in grayscale).}
	\label{fig:50kW}
\end{figure}

According to experimental observations, under low heat flux 20 kW/m$^2$, see Fig.\ref{fig:20kW}, two major cracks appeared: one circularly shaped and close to the sample edge, the other as a straight line right across the sample centre. The case with high heat flux 50 kW/m$^2$, shown in Fig.\ref{fig:50kW}, is more complicated, since the crack pattern can be divided into two zones. The cracks close to the sample edge presented a radial pattern, while those near the centre were distributed more randomly.


This is very interesting, as it shows that for this circular symmetry two different heat fluxes provide two distinct crack patterns on MDF. In \cite{Baroudi2017206} we observed something analogous in the case of square samples, where two distinct geometries are determined by two different materials (orthotropic fir wood and isotropic MDF). However, Figs.\ref{fig:20kW} and \ref{fig:50kW} show that the topology induced by circular geometry is very sensitive to the penetration depth, which is related to the heat flux magnitude as we explain later on in Section \ref{sec:experiments-interpretation-by-the-thermomechanical-model}.



\section{A full 3-D model of thermomechanical buckling and its validation}\label{sec:a-full-3-D-model-of-thermomechanical-buckling-and-its-validation}

In this section we describe the full 3-D model used to investigate the physical problem of MDF surface wrinkling.
Consider a three dimensional hot layer, subjected to in-plane thermal elongations that are restrained by an elastic, colder substrate to which it is perfectly bonded.
The according thermoelasticity formulation is based on the equilibrium equations
\be\label{equilibriumeq}
{\rm div} {\boldsymbol\sigma} + \rho {\bf f} = {\bf 0}\,,
\ee
with ${\boldsymbol\sigma}$ the stress tensor, $\rho$ its density and ${\bf f}$ the resultant of external forces. We label by ${\bf D}$ the symmetric elasticity tensor and by $\boldsymbol{\epsilon}^{(\rm th)}$ the thermal strain tensor. The general Hook's law then gives for the stress tensor of the material the following expression,
\be\label{hook}
\boldsymbol\sigma =  {\bf D}: \left(\boldsymbol\epsilon - \boldsymbol{\epsilon}^{(\rm th)}\right) \,.
\ee
The total deformation is thus written as
\be\label{straindisplacement}
\boldsymbol\epsilon = {1 \over 2} [(\nabla {\bf u})^{\rm T}  + \nabla {\bf u} +   (\nabla {\bf u})^{\rm T} \nabla {\bf u} ]\,,
\ee
where ${\bf u}$ is the displacement. The above deformation (or strain tensor) can be also rewritten as the direct sum of thermal and elastic strains $\boldsymbol\epsilon^{(\rm e)}$,
\be\label{totaldeformation}
\boldsymbol\epsilon = \boldsymbol\epsilon^{(\rm th)} + \boldsymbol\epsilon^{(\rm e)}
=
\boldsymbol\alpha  \Delta T + \boldsymbol\epsilon^{(\rm e)}\,.
\ee
$\boldsymbol\alpha$ is the thermal expansion tensor and $\Delta T$ is the temperature change. The MDF properties are derived from \cite{Sebera2014}.

The thermally induced displacement field ${\bf u}$ in principle has components $(u,v,w)$ along the axes $(x,y,z)$, but one can exploit the $U(1)$ symmetry of the circular sample and perform the substitution $(x,y,z)\mapsto (r,\theta,z)$, where $\theta \in [0,2\pi]$.
Eq.(\ref{equilibriumeq}) can be solved using the appropriate boundary conditions for a plate with free edge laying on an elastic foundation, namely null Kirchhoff effective shear force $V_r$ and zero radial bending moment $M_r$ (see also \ref{sec:approximate-2-D-analytical-solution} for the 2-D counterpart).
Our formulation considers only the action of thermal stresses, therefore in the equilibrium equation (\ref{equilibriumeq}) we set the resultant of external forces to zero, namely ${\bf f}={\bf 0}$.

We thus obtain an eigenvalue problem, which provides the most stressed locations of the hot surface\footnote{at such locations the major tensile stress is maximal, as demonstrated in \cite{Baroudi2017206}.}: these correspond indeed to the nodes of the resulting eigenmodes. As soon as the mechanical resistance of the material decreases with increasing temperature, the cracks will initiate along these nodes.


\begin{figure}

	\centering
	\includegraphics[width=0.48\textwidth]{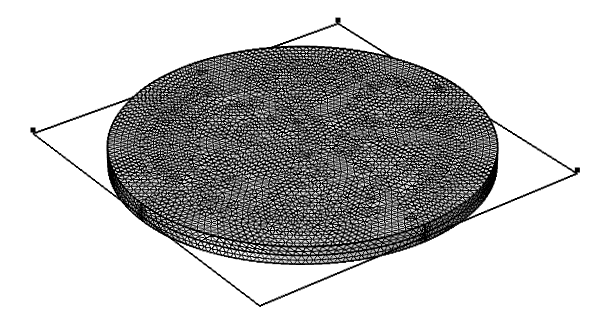}
	\caption{FEM mesh for the 3-D COMSOL simulations.}
	\label{fig:mesh}

\end{figure}

The model validation is done by comparing its predictions with a well-known 2-D analytical solution for the buckling of a steel disk, computed in \cite{Wang2004}.
Using the Poisson ratio for steel, namely $\nu\sim0.3$, we solve the equilibrium equations (\ref{equilibriumeq}) numerically by Finite Element Method (FEM) with the program COMSOL Multiphysics \cite{Pryor:2009:MMU:1823119}. The according tetrahedral volume mesh is shown in Figure \ref{fig:mesh}, and is composed of 3-D solid elements with quadratic discretization for the displacements.
 %
 
The result is plotted in Fig.\ref{fig:results}
%
The agreement is very good, the only difference being the larger eigenvalues of the 2-D solution \cite{Wang2004}, which is a dimensional reduction of the 3-D problem.
This is expected, as it is well known that the smallest eigenvalue $\lcr$ is enhanced at lower dimensions \cite{timoshenko2012theory}.

As discussed into detail in the Appendix, we formulate our thermomechanical model also in 2-D, which is solved both analytically and numerically.
Our 2-D curve is successfully validated, as it overlaps with the one found in the literature \cite{Wang2004} for any value of the relative stiffness parameter $\gamma\equiv R(k/D)^{1/4}$. Here $D\approx Eh_c^3/[12(1-\nu^2)]$ is the flexural rigidity, $E$ the Young modulus, $R$ and $h_c$ respectively the hot plate radius and thickness. The spring coefficient $k$ [N/m$^3$] is obtained by integrating the Boussinesq’s solution for this problem over a circle with unit radius. More details about $E$ and $D$, together with their numerical estimates, are given in \cite{Baroudi2017206}.

 It is easily seen that for low values of $\gamma$, our 2-D and 3-D solutions coincide, while when $\gamma\gtrsim 5$ the 3-D solution returns slightly lower values for $\lcr$. Moreover, compared to the analytical solution given for a steel disk, it seems that the numerical calculation finds additional modes for $n=2,3$. This is shown in Fig.\ref{fig:Nu_0p3_Simply_Supported} in \ref{sec:variational-formulation-and-isogeometric-analysis-of-a-2-D-stability-problem}.


\section{Interpretation of experiments by the thermomechanical model}\label{sec:experiments-interpretation-by-the-thermomechanical-model}

In this section we apply our thermomechanical model to the specific case of a round MDF sample, in order to verify that it is indeed able to reproduce the crack patterns observed in the experiments, shown in Figs.\ref{fig:20kW} and \ref{fig:50kW}.

To this aim, we substitute a Poisson ratio valid for wood $\nu=0.02$ above 150 \textdegree C \cite{Woodhandbook} in our 3-D and 2-D solutions, to compute the corresponding critical eigenmodes. According to several studies, the Poisson ratio for MDF is indeed of order $\mathcal{O}$(0.01-0.1); in particular, the authors of \cite{Sebera2014} have measured $\nu$=0.018-0.105 for an 18 mm thickness.
	As observed in \cite{Baroudi2017206}, by expanding the flexural rigidity $D\approx Eh_c^3/[12(1-\nu^2)]$ in powers of $\nu$ one can easily verify that, for the range above, the specific value of the Poisson ratio does not influence the location of global modes.

 Specifically, we notice the upper bound on $\lcr$ given by the 2-D solution, as already discussed. For several ranges of $\gamma$ we verify the accumulation of modes, reflecting the well-known fact that the system is very sensitive to perturbations, see for instance \cite{van2009w}. This pattern is found also in the observations.

\begin{figure}[t]
	\centering
	\includegraphics[width=0.49\textwidth]{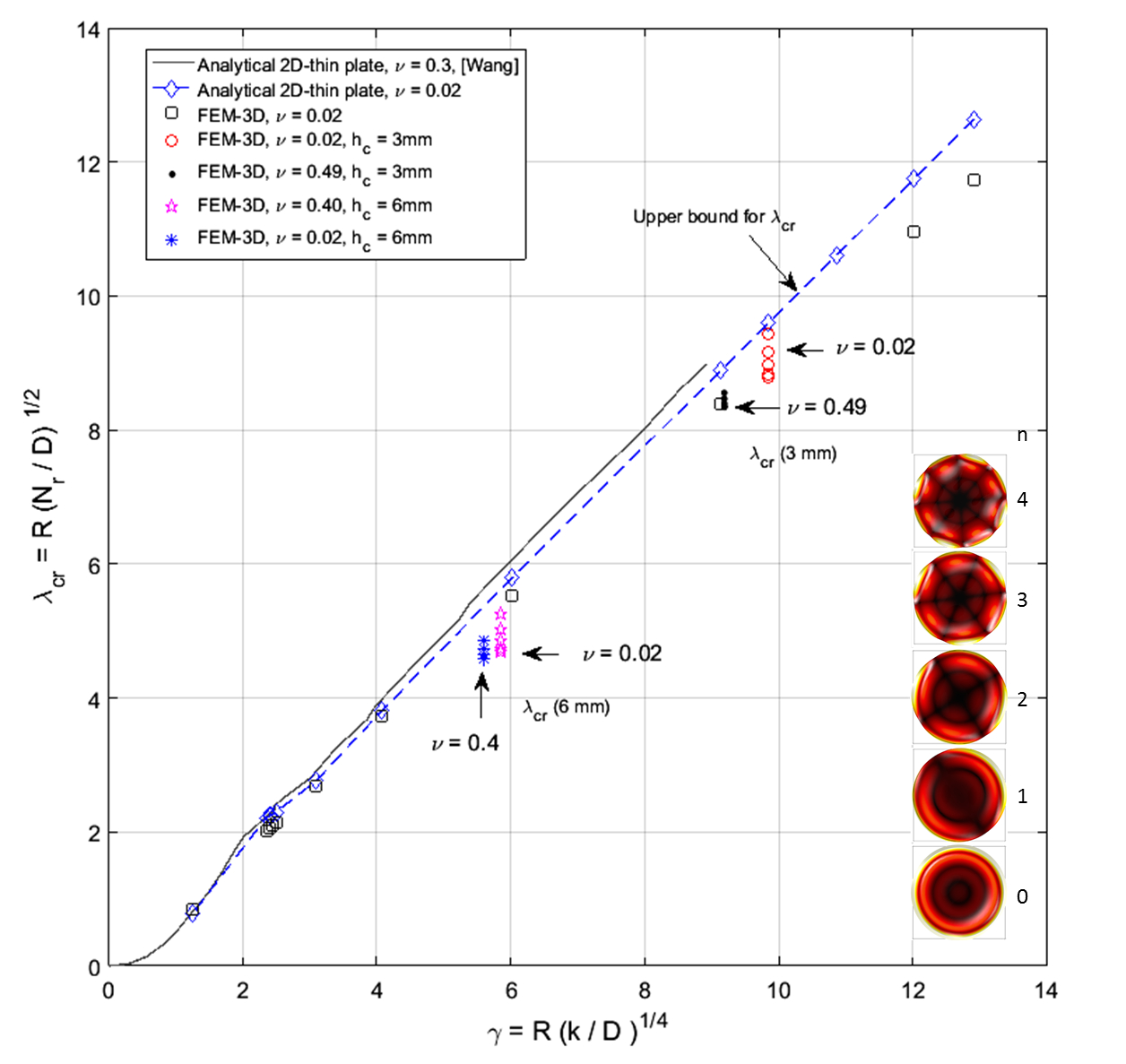}
	\caption{Model predictions and comparison with measurements.}
	\label{fig:results}
\end{figure}

Furthermore, one can see that when the ratio of foundation stiffness vs bending rigidity $\gamma$ is low, the buckling mode tends to be more global (zeroth and first mode on the right side of Fig.\ref{fig:results}). On the contrary, when the spring constant $k$ is large compared to the bending stiffness, we are in the presence of surface wrinkling (for instance 4th and higher modes).
%
A thicker heated layer therefore tends to exhibit global buckling modes (long wave lengths), while for a thinner one the surface wrinkles (short wave lengths).
Phenomenologically, this corresponds respectively to Fig.\ref{fig:20kW}, i.e. low heat flux 20 W/m$^2$, and to Fig.\ref{fig:50kW}, high heat flux 50 W/m$^2$. The penetration depth is indeed inversely proportional to the incident heat flux: when this is low, it takes longer to the surface to reach the same temperature, and the heat can travel deeper inside the material. This leads to a longer penetration depth \cite{a3f4a276a46c4ecaa357a2fd51e95882}.

Notice also how the 3-D model can fully explain the vertical crack on the specimen side, along the thickness (Figure \ref{fig:results}). Such a crack usually appears along the direction of the principal plane. Finally, the additional patterns that appear in the central area of the specimens (see e.g. Fig.\ref{fig:50kW}) can be probably explained by secondary bifurcations. This is evident from the Y-shaped cracks (or sulci, see \cite{PhysRevLett.110.024302}) within the "bubbles" shown in Fig.\ref{fig:mdfpatterns}. However, a rigorous study of these patterns requires a full non-linear analysis that goes beyond the scope of this article.

\section{Conclusions}\label{sec:conclusions}
 
In this paper we have considered the formation of crack patterns on the surface of circular MDF samples subjected to radiation heat flux in inert atmosphere. Our analysis shows that a model of thermomechanical surface instability is able to reproduce the observed cracks pattern topology, while providing at the same time a possible physical explanation for their formation.

Interestingly, the crack patterns for both circular and square MDF specimens are recreated by the same model of surface instability for a plate over an elastic foundation: the macroscopic physical mechanism is indeed identical.
Furthermore, the same effect was observed in \cite{Lyukshin2018} for the case of a thermal barrier (oxide ceramic) coating, buckled on an elastic substrate under thermal shock. The authors could successfully explain the buckling modes via a model that is formally analogous to ours. This seems to suggest that the physical phenomenon here investigated is rather general, applying to other plastic materials as well as wood, which is a natural thermoplastic.

It is important to note that the present study is however inherently \textit{qualitative}, meaning that we are mostly concerned about verifying whether the same thermomechanical instability mechanism applies to circular as well as to square MDF samples.
	This paper thus consolidates the qualitative value of the phenomenological model proposed in \cite{Baroudi2017206} for the crack-pattern formation. There is no aim in computing exactly where and how much a crack opens:
we will address the quantitative predictions in terms of the crack patterns geometry in a forthcoming paper.

We should also remark that our explanation for the crack patterns is valid for temperatures which are lower than those which induce chemical decomposition\footnote{We anyway expect that, under our conditions, the eventual presence of oxygen or moisture would not change these results appreciably, if not at all.} (pyrolysis point $T_p\approx 300 ^\circ{\mathrm C}$). Also, we do not address combustion in the gas phase (flaming).

If one aims to explore what happens at higher temperatures, the emerging chemico-physical phenomena concur in creating a much more complex phenomenology.
For instance, it is well known (see e.g. \cite{ROBERTS1971893}) that once the solid materials like cellulose, rubber, and plastics are burned, they release combustible gases via pyrolysis, under the effects of external and flame heat fluxes. The materials would lose their structural integrity by charring, deforming and developing defects such as cracks, bubbles and voids. For wooden materials, the char shrinkage and cracking are typical "charring behaviours", which reduce the heat barrier effect of the char layer during flaming combustion and pyrolysis. These defects enhance the combustion process by allowing oxygen and external heat flux to travel further into the material; they also allow pyrolysis gases to escape to the surface for subsequent combustion.


The implications of this work in terms of pyrolysis modelling of wood are thus evident.
 Unfortunately, no quantitative assessment has been performed so far, however ongoing experiments seem to suggest that the heat transfer enhancement affects materials mainly near the surface, with additional formation of fissure flames. We accordingly believe that our efforts in describing the fissures formation will eventually have an impact on our understanding of pyrolysis modelling.

Finally, to refine and fully validate our model, further experiments on both MDF and wooden samples will be needed. These should include at least measurements of temperature profile and deformations (displacements and strains), for capturing the transition to surface instability. Also, in order to identify precisely the relevant temperature range one needs the experimental determination of the glass-transition point. Measurements of thermal expansion coefficient and elasticity modulus in function of temperature at the macroscopic scale are also required.

Nevertheless, even taking into account these limitations, the topology of cracks  observed on MDF samples seem to be well explained by a thermomechanically driven surface instability that occurs before pyrolysis.
Specifically, in this paper we have confirmed that such phenomenon is totally general, independently of the particular shape of the specimens and of the different crack patterns.

\section*{Aknowledgements}

This work was supported by the National Natural Science Foundation of China (NSFC) under Grant No. 51876148. AF acknowledges the Estonian Research Council with Institutional research funding grant IUT1-15, and the Estonian Centre of Excellence in Zero Energy and Resource Efficient Smart Buildings and Districts, ZEBE, grant 2014-2020.4.01.15-0016 funded by the European Regional Development Fund.

SK has been supported by the Academy of Finland through the project 
{\it Adaptive isogeometric methods for thin-walled structures} 
(decision numbers 270007, 273609, 304122). Access and licenses for the commercial software Abaqus FEA have been provided by CSC-IT Center for Science.

\appendix

\section{Approximate 2-D analytical solution for a thin plate}\label{sec:approximate-2-D-analytical-solution}

In this Appendix we show that the full 3-D numerical model described in Section \ref{sec:a-full-3-D-model-of-thermomechanical-buckling-and-its-validation} can be successfully reduced to an analytical model in 2-D, by assuming a circular soft thin plate bonded to an elastic foundation. The corresponding eigenvalue problem is solved both analytically and numerically.

The governing equation for a thin plate bounded by an elastic Winkler foundation is generally written as (see \cite{Baroudi2017206} and references quoted therein)
\be\label{bucklingeq}
\nabla^4 w+\lambda^2 \nabla^2w+\gamma^4w=0\,,
\ee
for the displacement field $w$, with lengths normalized by the plate radius $R$. We treated the case of rectangular plate in \cite{Baroudi2017206}, where $\nabla$ is the gradient in Cartesian coordinates. For a circular plate instead, the load parameter is $\lambda^2\equiv R^2 N_r/D$, and $N_r$ is the uniform radial load at the edge.
$D\approx Eh_c^3/[12(1-\nu^2)]$ is the flexural rigidity, with the Young modulus $E$ and the hot plate thickness $h_c$.
$\gamma^4\equiv kR^4/D$ is a stiffness parameter, where $k$ is the spring constant of the foundation (modelled as described into detail in \cite{Baroudi2017206}, where $\gamma\equiv\beta$). For a radial symmetry, assuming $n\in{\mathrm N}$ nodal diameters, the solution of the buckling equation (\ref{bucklingeq}) can be written in polar coordinates ($r,\theta$) as
\be
w=u(r)\cos(n\theta)\,, \qquad \theta\in[0,2\pi]\,,
\ee
so that Eq.(\ref{bucklingeq}) can be recast as follows,
\be\label{bucklingeqU1}
L^2u+\lambda^2Lu+\gamma^4 u =0\,,
\ee
with the Laplacian operator in polar coordinates
\be
L\equiv \dfrac{d^2}{dr^2}+\dfrac{1}{r}\dfrac{d}{dr}-\dfrac{n^2}{r^2}\,.
\ee
The eigenvalue problem for our case is given by Eq.(\ref{bucklingeqU1}) and the following boundary conditions for a plate with free edge on an elastic foundation \cite{Wang2004},
\begin{align}
\label{bc1}
&u^{\prime\prime}(1)+\nu[u^\prime(1)-n^2u(1)]=0\,,
\\
&u^{\prime\prime\prime}(1)+u^{\prime\prime}(1)-[1+n^2(2-\nu)-\lambda^2]u^\prime(1)
\nonumber\\
&+n^2(3-\nu)u(1)=0\,,
\label{bc2}
\end{align}
namely zero moment and resultant shear, respectively.

Depending on the relative magnitude of $\gamma$ and $\lambda$, we obtain three general different solutions which are bounded at the origin. If $J_n$ is the Bessel function of the first kind of order $n$, for $\lambda>\rad\gamma$ we obtain the classical solution by \cite{Wang2004},
\be
u(r)=C_1J_n(\alpha r)+C_2J_n(\beta r)\,,
\ee
where
\be
\alpha=\sqrt{\dfrac{\lambda^2+\sqrt{\lambda^4-4\gamma^4}}{2}}\,,
\;
\beta=\sqrt{\dfrac{\lambda^2-\sqrt{\lambda^4-4\gamma^4}}{2}}\,.
\ee
Substituting in the b.c. (\ref{bc1}) and (\ref{bc2}), one finds the following criticality condition,
\begin{align}
\label{crit1}
&f(\lambda)=
[\alpha^2J_n^{\prime\prime}(\alpha r)+\nu(\alpha J^\prime_n(\alpha r)-n^2J_n(\alpha r))]
\nonumber\\
&\times[
\beta^3J_n^{\prime\prime\prime}(\beta r)+\beta^2J_n^{\prime\prime}(\beta r)
-[1+n^2(2-\nu)-\lambda^2]\nonumber\\
&
\times\beta J_n^{\prime}(\beta r)+n^2(3-\nu)J_n(\beta r)
]
\nonumber\\
&-
[\beta^2J_n^{\prime\prime}(\beta r)+\nu(\beta J^\prime_n(\beta r)-n^2J_n(\beta r))]
\nonumber\\
&\times[
\alpha^3J_n^{\prime\prime\prime}(\alpha r)+\alpha^2J_n^{\prime\prime}(\alpha r)
-[1+n^2(2-\nu)-\lambda^2]\nonumber\\
&\times
\alpha J_n^{\prime}(\alpha r)+n^2(3-\nu)J_n(\alpha r)
]=0\,.
\end{align}
The critical $\lcr$ is the lowest value of $\lambda$ satisfying the above.

\begin{figure}[t]
	
	
	\centering
	\includegraphics[width=0.49\textwidth]{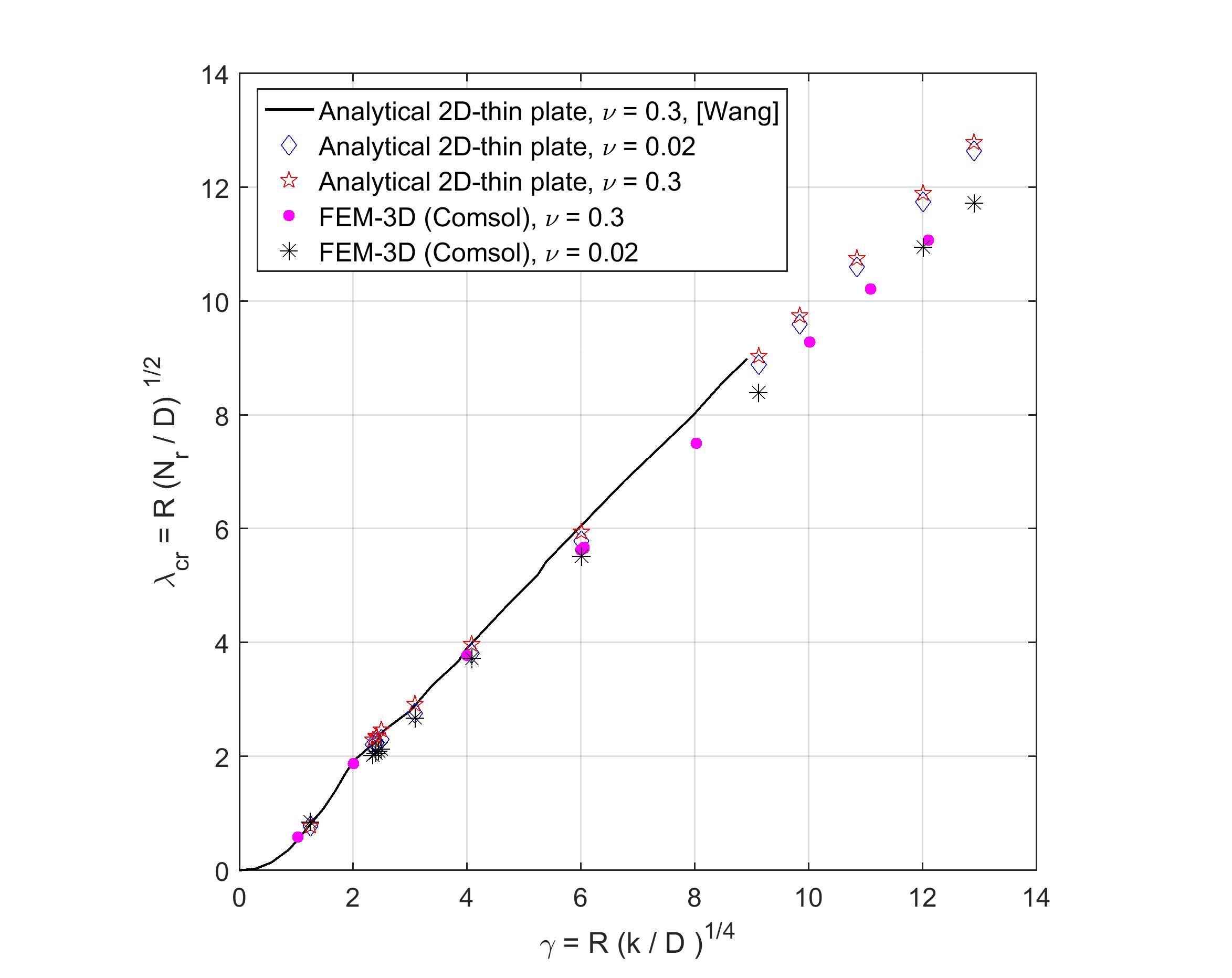}
	\caption{Validation of the model (for $\nu=0.3$) and both 3-D and 2-D prediction for $\nu=0.02$.}
	\label{fig:validation}
\end{figure}

For $\lambda=\rad\gamma$, the solution is instead
\be
u=C_1J_n(\gamma r)+C_2rJ_{n+1}(\gamma r)\,,
\ee
which gives the criticality condition
\begin{align}
\label{crit2}
&f(\lambda)=[\gamma^2J_n^{\prime\prime}(\gamma r)+\nu(\gamma J_n^{\prime}(\gamma r)- n^2 J_n(\gamma r))]
\nonumber\\
&\times\{
3\gamma^2J_{n+1}^{\prime\prime}(\gamma r)+\gamma^3rJ_{n+1}^{\prime\prime\prime}(\gamma r)
+2\gamma J_{n+1}^{\prime}(\gamma r)
\nonumber\\
&
+\gamma^2rJ_{n+1}^{\prime\prime}(\gamma r)
-[1+n^2(2-\nu)-\lambda^2]
\nonumber\\
&
\times[J_{n+1}(\gamma r)+\gamma r J_{n+1}^{\prime}(\gamma r)]+n^2(3-\nu)rJ_{n+1}(\gamma r)
\}
\nonumber\\
&-
\{\gamma^3J_{n}^{\prime\prime\prime}(\gamma r)+\gamma^2J_{n}^{\prime\prime}(\gamma r)
-[1+n^2(2-\nu)-\lambda^2]
\nonumber\\
&
\times
\gamma J_{n}^\prime(\gamma r)+n^2(3-\nu)J_{n}(\gamma r)
\}
\nonumber\\
&\times
[2\gamma J_{n+1}^{\prime}(\gamma r)+\gamma^2 r J_{n+1}^{\prime\prime}(\gamma r)
+\nu J_{n+1}(\gamma r)
\nonumber\\
&
+\gamma r J_{n+1}^{\prime}(\gamma r)
-n^2 rJ_{n+1}(\gamma r)
]=0\,.
\end{align}
Finally, for $\lambda<\rad\gamma$, the solution is written as
\be
u=C_1\mathrm{Re}[J_n(i\delta r)]+C_2\mathrm{Im}[J_n(i\delta r)]\,,
\ee
where $i^2=-1$ and
\be
\delta=\sqrt{\dfrac{-\lambda^2+i\sqrt{4\gamma^4-\lambda^4}}{2}}\,.
\ee
In this case the criticality condition holds as follows,
\begin{align}\label{crit3}
&f(\lambda)=\{\mathrm{Re}^{\prime\prime}[J_n]+\nu(\mathrm{Re}^{\prime}[J_n]-n^2\mathrm{Re}[J_n])\}
\nonumber\\
&\times
\{\mathrm{Im}^{\prime\prime\prime}[J_n]+\mathrm{Im}^{\prime\prime}[J_n]
-[1+n^2(2-\nu)-\lambda^2]\mathrm{Im}^{\prime}[J_n]
\nonumber\\
&+n^2(3-\nu)\mathrm{Im}[J_n]
\}
-(\mathrm{Re}\Leftrightarrow\mathrm{Im})=0\,.
\end{align}
Solving Eqs.(\ref{crit1}), (\ref{crit2}) and (\ref{crit3}) returns the values of $\lcr$ that are plotted in Figs.\ref{fig:results} and \ref{fig:validation}.

\section{Variational formulation and isogeometric analysis of a 2-D stability problem}\label{sec:variational-formulation-and-isogeometric-analysis-of-a-2-D-stability-problem}

The weak (or variational) formulation of the stability problem
for a plate on elastic foundation corresponding to \eqref{bucklingeqU1}
with
\eqref{bc1}
and
\eqref{bc2}
reads as follows. Find
$ u \in U \subset H^2(\Omega) $
such that
\begin{equation}
\label{weak_1}
a(u, v) = \lambda^2 m(u, v)
\quad
\forall v \in V \subset H^2(\Omega) ,
\quad
\Omega = (0,1)
\end{equation}
where the bilinear forms $a$ and $m$:
$ U \times V \rightarrow \mathbb{R} $,
respectively, are defined as
\begin{align}
\label{a}
&a(u, v)
=
(\nu - 1) 
 \displaystyle\int_{0}^{1}
\left[ u'' \left( \dfrac{v'}{r} - n^2 \dfrac{v}{r^2} \right)
\right.&\nonumber
\\
&+\left.
 v'' \left( \dfrac{u'}{r} - n^2 \dfrac{u}{r^2} \right)- 2 n^2 \left( \dfrac{u'}{r} - \dfrac{u}{r^2} \right)
\left( \dfrac{v'}{r} - \dfrac{v}{r^2} \right) \right] r dr
 &\nonumber
\\
&+ \displaystyle\int_{0}^{1}
\left( u'' + \frac{u'}{r} - n^2 \frac{u}{r^2} \right)
\left( v'' + \frac{v'}{r} - n^2 \frac{v}{r^2} \right) r dr
\nonumber\\
&+ \gamma^4
\displaystyle\int_{0}^{1} u v r dr,
\end{align}
and
\be
\label{m}
m(u, v) =
\int\limits_{0}^{1}
\left( u'v' + n^2 \frac{u}{r} \frac{v}{r} \right) r dr ,
\ee
with $ u(r) $ and $ v(r) $
standing, respectively, for trial and test functions, where the prime denotes differentiation with respect to the radial coordinate $r$.
The homogeneous Neumann boundary conditions
at free edge $ r = 1 $
are fulfilled automatically.
In the corresponding conforming Galerkin formulation, one finds
$ u_h \in U_h \subset U $
such that
\be
\label{weak_2}
a(u_h, v_h) = \lambda^2 m(u_h, v_h)
\quad
\forall v_h \in V_h \subset V .
\ee
An isogeometric NURBS-based discretization of the solution domain
naturally provides
$ C^{p-1} $ global regularity
\cite{hughes-cottrell-basilevs-05},
where $p$ is a B-spline order.
For $ p \geq 2 $,
the corresponding isoparametric discrete function space
is a subset of an $H^2$ Sobolev space,
which provides a conforming Galerkin version of the method.

\begin{figure}[t]
		\centering
		\begin{minipage}{0.45\textwidth}
			\centering
			\includegraphics[width=\textwidth]{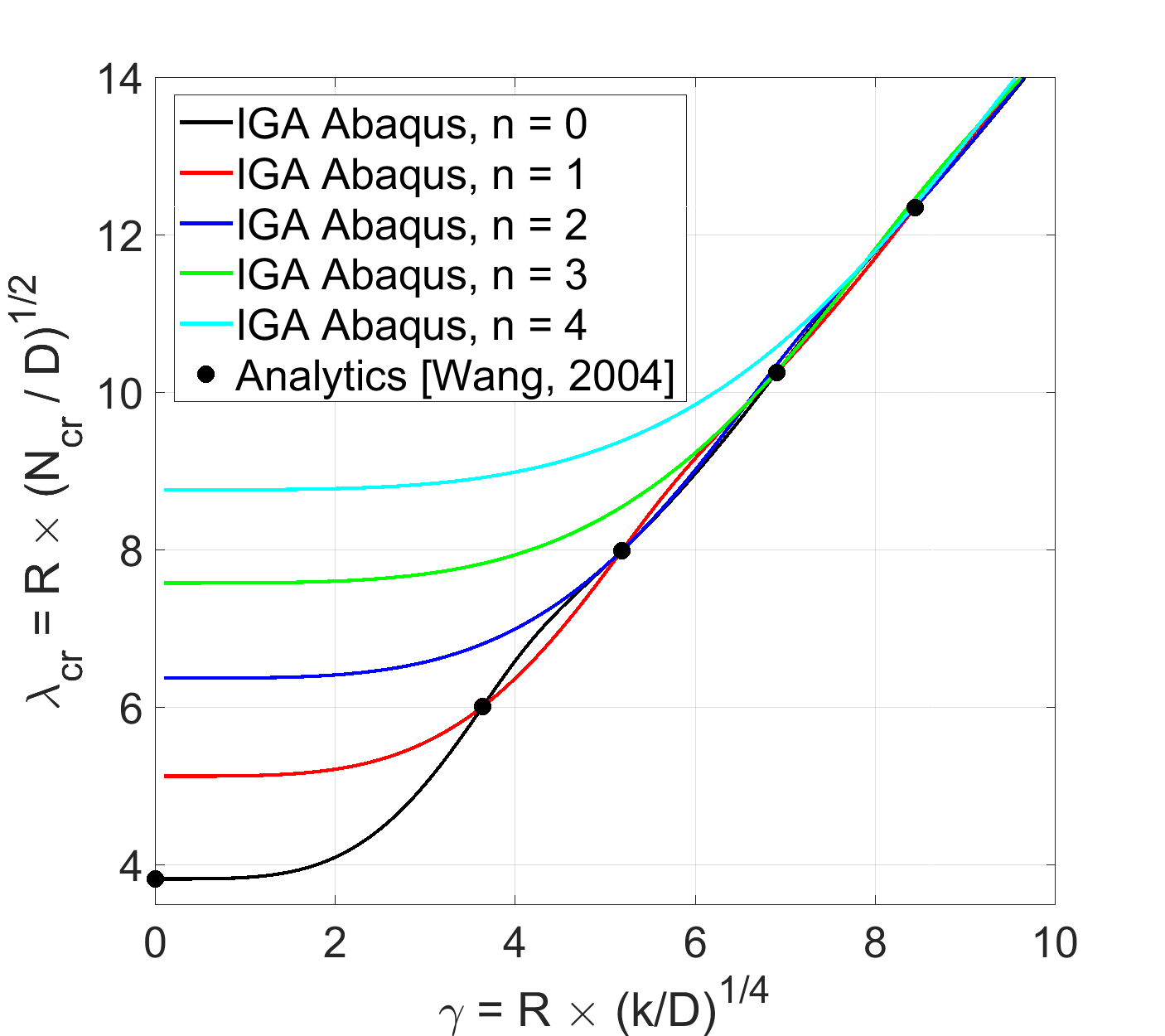}
			\caption{$\nu = 0.3 $, Clamped edge }
			\label{fig:Nu_0p3_Clamped}
	\end{minipage}
		\begin{minipage}{0.45\textwidth}
	\centering
			\includegraphics[width=\textwidth]{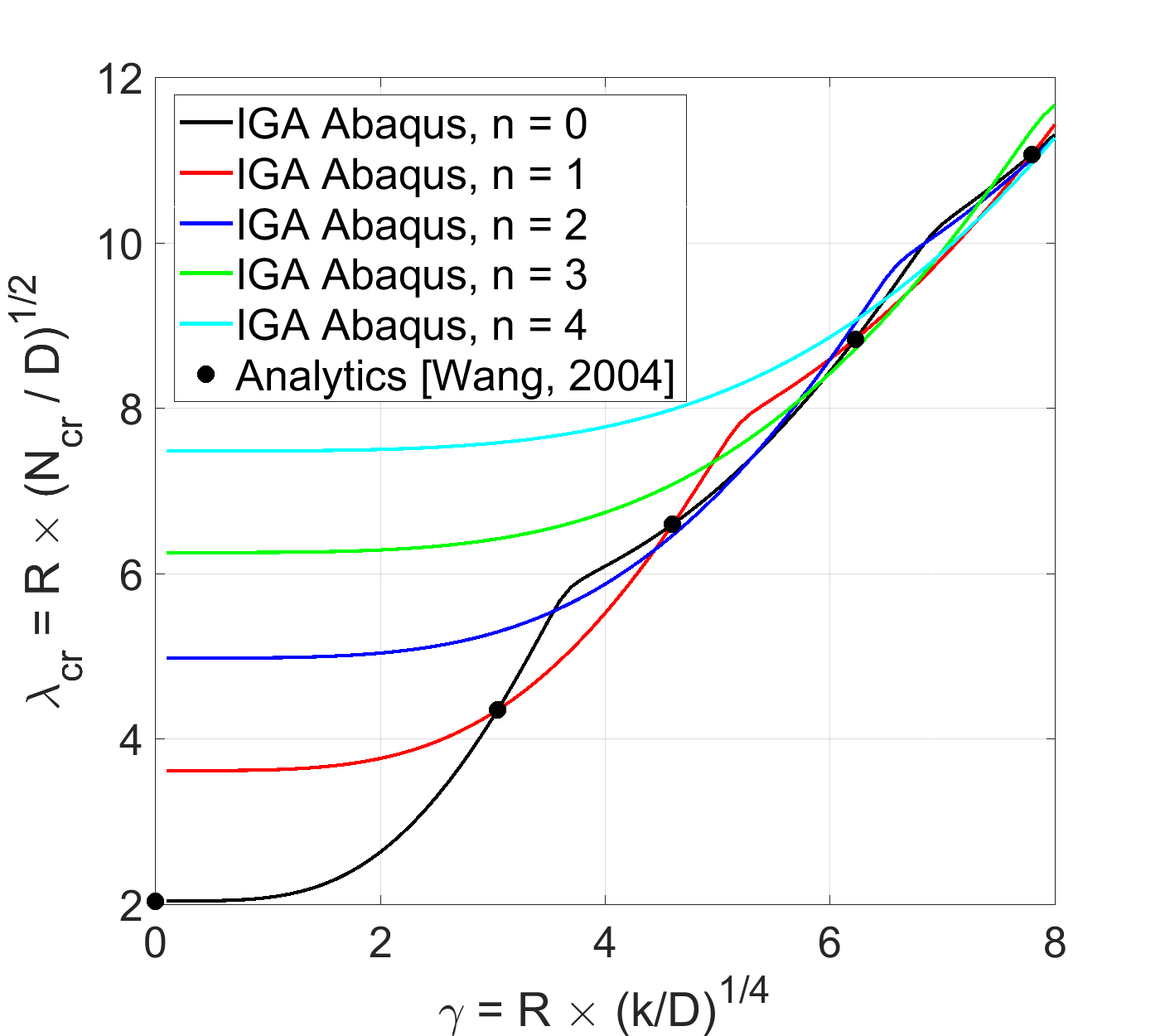}
			\caption{ $\nu = 0.3 $, Simply supported edge }
			\label{fig:Nu_0p3_Simply_Supported}
	\end{minipage}

\end{figure}

\begin{figure}[t]
	\centering
		\begin{minipage}{0.45\textwidth}
			\includegraphics[width=\textwidth]{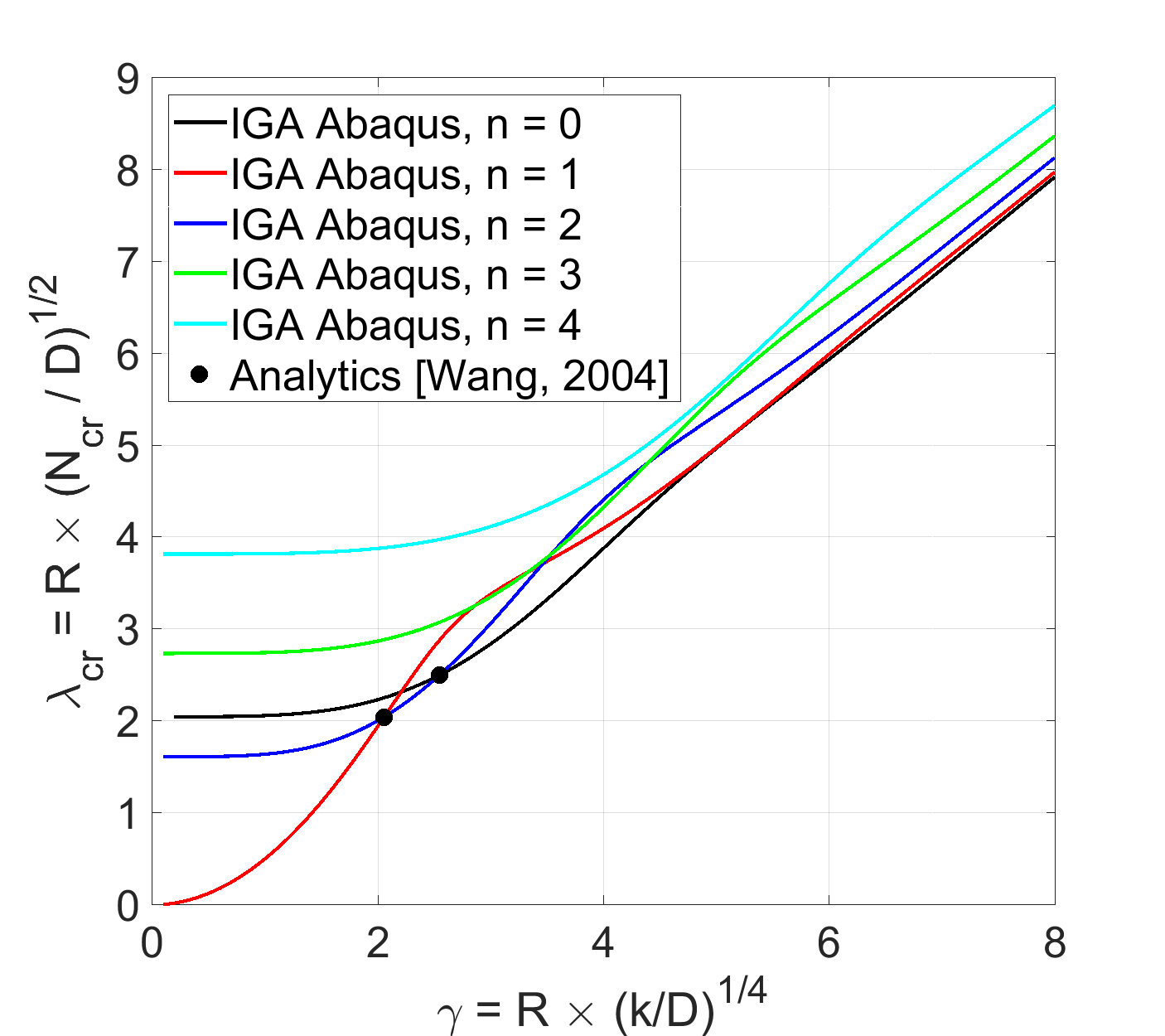}
			\caption{ $\nu = 0.3 $, Free edge }
			\label{fig:Nu_0p3_Free_Edge}
	\end{minipage}
		\begin{minipage}{0.45\textwidth}
		\centering
			\includegraphics[width=\textwidth]{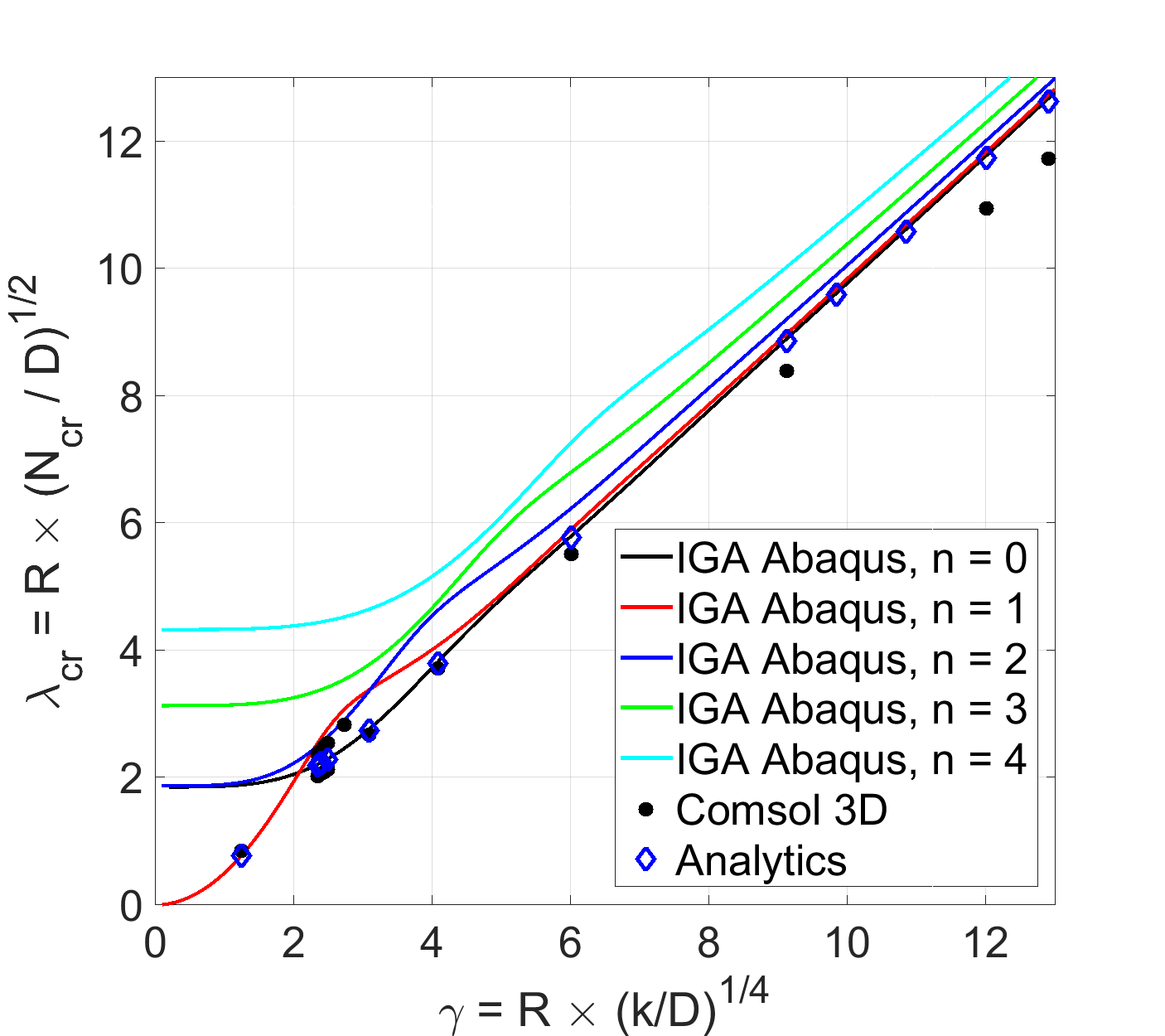}
			\caption{ $\nu = 0.02 $, Free edge }
			\label{fig:Nu_0p02_Free_Edge}
	\end{minipage}

	\end{figure}

The numerical implementation utilizing user-defined
finite elements of the commercial software Abaqus FEA
is described in
\cite{khakalo-niiranen-2017}.
For steel plates with $ \nu = 0.3 $,
we consider three types of boundary conditions corresponding to
clamped, simply supported and free edges.
The critical load $ \lcr $ against the foundation stiffness $ \gamma $ is presented for different cases of radial symmetries ($ n = 0, ... , 4 $) in Figs.\ref{fig:Nu_0p3_Clamped},
\ref{fig:Nu_0p3_Simply_Supported} and
\ref{fig:Nu_0p3_Free_Edge}, respectively.
Solid curves represent the values calculated via Abaqus user elements.
Circle marks, with values analytically defined in
\cite{Wang2004},
are used for the verification of the numerical implementation.
The lowest curves build the border line of critical buckling loads.
For the simply supported case,
it should be mentioned that the border line of $ \lambda_{cr} $
is composed of several curves corresponding to $ n = 0, 1, ... $,
while in
\cite{Wang2004}
only two curves ($ n = 0 $ and $ n = 1 $) define the border line.

The case when $ \nu = 0.02 $, which concerns the MDF samples in our experiments, is shown in Fig.\ref{fig:Nu_0p02_Free_Edge}.
Diamond marks correspond to analytical values, while circle marks stand for those calculated with the FEM software COMSOL.

It is worth noting that, as a side result, our numerical 2-D analysis finds additional modes (for $n=2,3$) compared to the analytical solutions given in \cite{Wang2004} for a steel disk, Figure \ref{fig:Nu_0p3_Free_Edge}.

%
%
%

%

\newpage

\section*{References}

\bibliography{thermo-mechanical_circle}

\begin{thebibliography}{10}
\expandafter\ifx\csname url\endcsname\relax
  \def\url#1{\texttt{#1}}\fi
\expandafter\ifx\csname urlprefix\endcsname\relax\def\urlprefix{URL }\fi
\expandafter\ifx\csname href\endcsname\relax
  \def\href#1#2{#2} \def\path#1{#1}\fi

\bibitem{ROBERTS1971893}
A.~Roberts, Problems associated with the theoretical analysis of the burning of
  wood, Symposium (International) on Combustion 13~(1) (1971) 893 -- 903,
  thirteenth symposium (International) on Combustion.
\newblock \href
  {http://dx.doi.org/http://dx.doi.org/10.1016/S0082-0784(71)80090-5}
  {\path{doi:http://dx.doi.org/10.1016/S0082-0784(71)80090-5}}.

\bibitem{Babrauskas2005528}
V.~Babrauskas, Charring rate of wood as a tool for fire investigations, Fire
  Safety Journal 40~(6) (2005) 528 -- 554.
\newblock \href
  {http://dx.doi.org/http://dx.doi.org/10.1016/j.firesaf.2005.05.006}
  {\path{doi:http://dx.doi.org/10.1016/j.firesaf.2005.05.006}}.

\bibitem{Ross}
R.~J. Ross, R.~H. White, Wood Condition Assessment Manual: Second Edition,
  {FPL} ; {GTR}-234 Edition, Forest Products Society, 2014, {G}eneral technical
  report. Post-Fire Assessment of Structural Wood Members, pp. 29--46.

\bibitem{doi:10.1080/10618562.2012.659663}
K.~McGrattan, R.~McDermott, J.~Floyd, S.~Hostikka, G.~Forney, H.~Baum,
  Computational fluid dynamics modelling of fire, International Journal of
  Computational Fluid Dynamics 26~(6-8) (2012) 349--361.
\newblock \href {http://dx.doi.org/10.1080/10618562.2012.659663}
  {\path{doi:10.1080/10618562.2012.659663}}.

\bibitem{Stoliarov20102024}
S.~I. Stoliarov, S.~Crowley, R.~N. Walters, R.~E. Lyon, Prediction of the
  burning rates of charring polymers, Combustion and Flame 157~(11) (2010) 2024
  -- 2034.
\newblock \href
  {http://dx.doi.org/http://dx.doi.org/10.1016/j.combustflame.2010.03.011}
  {\path{doi:http://dx.doi.org/10.1016/j.combustflame.2010.03.011}}.

\bibitem{Lautenberger20091503}
C.~Lautenberger, C.~Fernandez-Pello, A model for the oxidative pyrolysis of
  wood, Combustion and Flame 156~(8) (2009) 1503 -- 1513.
\newblock \href
  {http://dx.doi.org/http://dx.doi.org/10.1016/j.combustflame.2009.04.001}
  {\path{doi:http://dx.doi.org/10.1016/j.combustflame.2009.04.001}}.

\bibitem{Baroudi2017206}
D.~Baroudi, A.~Ferrantelli, K.~Y. Li, S.~Hostikka, A thermomechanical
  explanation for the topology of crack patterns observed on the surface of
  charred wood and particle fibreboard, Combustion and Flame 182 (2017) 206 --
  215.
\newblock \href
  {http://dx.doi.org/https://doi.org/10.1016/j.combustflame.2017.04.017}
  {\path{doi:https://doi.org/10.1016/j.combustflame.2017.04.017}}.

\bibitem{Salmen}
L.~Salm{\'e}n, Viscoelastic properties of in situ lignin under water-saturated
  conditions, Journal of Materials Science 19~(9) (1982) 3090--3096.
\newblock \href {http://dx.doi.org/10.1007/BF01026988}
  {\path{doi:10.1007/BF01026988}}.

\bibitem{Antoniow2012}
J.~S. Antoniow, J.~E. Maigret, C.~Jensen, N.~Trannoy, M.~Chirtoc, J.~Beaugrand,
  Glass-transition temperature profile measured in a wood cell wall using
  scanning thermal expansion microscope (sthem), International Journal of
  Thermophysics 33~(10) (2012) 2167--2172.
\newblock \href {http://dx.doi.org/10.1007/s10765-012-1313-y}
  {\path{doi:10.1007/s10765-012-1313-y}}.

\bibitem{Bazant_1985}
M.~M. Hassania, F.~K. Wittela, S.~Heringa, H.~J. Herrmanna, Constitutive
  equation of wood at variable hymidity and temperature, Wood Sci Technol 19
  (1985) 159--177.

\bibitem{Salmen_1984}
L.~Salmen, Micromechanical understanding of the cell wall structure, C. R.
  Biologies 327.

\bibitem{Wang2004}
C.~Y. Wang, On the buckling of a circular plate on an elastic foundation,
  Journal of Applied Mechanics 72~(5) (2004) 795--796.
\newblock \href {http://dx.doi.org/10.1115/1.1988347}
  {\path{doi:10.1115/1.1988347}}.

\bibitem{a3f4a276a46c4ecaa357a2fd51e95882}
K.~Li, S.~Hostikka, P.~Dai, Y.~Li, H.~Zhang, J.~Ji, Charring shrinkage and
  cracking of fir during pyrolysis in inert atmosphere and different ambient
  pressures, PROCEEDINGS OF THE COMBUSTION INSTITUTE 36~(2) (2017) 3185–3194.
\newblock \href {http://dx.doi.org/10.1016/j.proci.2016.07.001}
  {\path{doi:10.1016/j.proci.2016.07.001}}.

\bibitem{LI201539}
K.~Li, D.~S. Pau, J.~Wang, J.~Ji, Modelling pyrolysis of charring materials:
  determining flame heat flux using bench-scale experiments of medium density
  fibreboard (mdf), Chemical Engineering Science 123 (2015) 39 -- 48.
\newblock \href {http://dx.doi.org/http://dx.doi.org/10.1016/j.ces.2014.10.043}
  {\path{doi:http://dx.doi.org/10.1016/j.ces.2014.10.043}}.

\bibitem{mcgrattan2013fire}
K.~McGrattan, S.~Hostikka, R.~McDermott, J.~Floyd, C.~Weinschenk, K.~Overholt,
  Fire dynamics simulator technical reference guide volume 1: Mathematical
  model, NIST special publication 1018.

\bibitem{Sebera2014}
V.~Sebera, J.~Tippner, M.~{\v{S}}imek, J.~{\v{S}}rajer, D.~D{\v{e}}ck{\'y},
  H.~Kl{\'i}mov{\'a}, Poisson's ratio of the mdf in respect to vertical density
  profile, European Journal of Wood and Wood Products 72~(3) (2014) 407--410.
\newblock \href {http://dx.doi.org/10.1007/s00107-014-0780-1}
  {\path{doi:10.1007/s00107-014-0780-1}}.

\bibitem{Pryor:2009:MMU:1823119}
R.~W. Pryor, Multiphysics Modeling Using COMSOL: A First Principles Approach,
  1st Edition, Jones and Bartlett Publishers, Inc., USA, 2009.

\bibitem{timoshenko2012theory}
S.~Timoshenko, J.~Gere, Theory of Elastic Stability, Dover Civil and Mechanical
  Engineering, Dover Publications, 2012.

\bibitem{Woodhandbook}
D.~W. Green, J.~E. Winandy, D.~E. Kretschmann, Wood handbook : wood as an
  engineering material, {FPL} ; {GTR}-113 Edition, Madison, WI: USDA Forest
  Service, Forest Products Laboratory, 1999, {G}eneral technical report.
  Mechanical properties of wood, pp. 4.1--4.45.

\bibitem{van2009w}
A.~van~der Heijden, W.T. Koiter's Elastic Stability of Solids and Structures,
  Cambridge University Press, 2009.

\bibitem{PhysRevLett.110.024302}
T.~Tallinen, J.~S. Biggins, L.~Mahadevan, Surface sulci in squeezed soft
  solids, Phys. Rev. Lett. 110 (2013) 024302.
\newblock \href {http://dx.doi.org/10.1103/PhysRevLett.110.024302}
  {\path{doi:10.1103/PhysRevLett.110.024302}}.

\bibitem{Lyukshin2018}
P.~A. Lyukshin, B.~A. Lyukshin, N.~Y. Matolygina, S.~V. Panin, Stress-strain
  state in a buckled thermal barrier coating on an elastic substrate, Physical
  Mesomechanics 21~(6) (2018) 498--507.
\newblock \href {http://dx.doi.org/10.1134/S1029959918060048}
  {\path{doi:10.1134/S1029959918060048}}.

\bibitem{hughes-cottrell-basilevs-05}
T.~J.~R. Hughes, J.~A. Cottrell, Y.~Bazilevs, Isogeometric analysis: Cad,
  finite elements, nurbs, exact geometry and mesh refinement, Comput. Methods
  Appl. Mech. Engrg. 194 (2005) 4135--4195.

\bibitem{khakalo-niiranen-2017}
S.~Khakalo, J.~Niiranen, Isogeometric analysis of higher-order gradient
  elasticity by user elements of a commercial finite element software,
  Computer-Aided Design 82 (2017) 154--169.

\end{thebibliography}

\end{document}